\documentclass[onecolumn,authoryear]{els-mrw} 

\usepackage{amsmath,amssymb,amsfonts,amsthm,makeidx,graphicx}
\usepackage{txfonts}
\usepackage{helvet}
\newcommand{\bea}{\begin{align}}
\newcommand{\eea}{\end{align}}

\newcommand{\Ryd}{\ensuremath{R_{\infty}}}
\def\vq{\vec q}

\def\a{\alpha}
\def\D{\Delta}\def\d{\delta}

\def\m{\mu}

\def\vq{\vec q}

\def\a{\alpha}
\def\p{\psi}
\newcommand{\rp}{\ensuremath{r_{\mathrm p}}}
\newcommand{\la}{\langle}
\newcommand{\ra}{\rangle}
\newcommand{\eq}[1]{Eq.~(\ref{#1})}
\newcommand{\mup}{\ensuremath{\mu \mathrm{p} }}\newcommand{\Htwo}{\ensuremath{\mathrm H_{2}}}
\newcommand{\Water}{\ensuremath{\mathrm H_{2}\mathrm O}}
\newcommand{\mus}{\ensuremath{\mu \mathrm{s} }}
\newcommand{\mum}{\ensuremath{\mu \mathrm{m} }}
\newcommand{\Ka}{\ensuremath{\mathrm{K}_\alpha }}

\begin{document}

\chapter{The Proton Radius Puzzle }\label{chap1}

\author[1]{Gerald A. Miller}%


\address[1]{\orgname{University of Washington}, \orgdiv{Physics  Department}, \orgaddress{Seattle, Washington 98195-1560, USA}}

\articletag{Chapter Encyclopedia of Nuclear Physics.}

\maketitle

\begin{glossary}[Glossary]
\term{Lamb shift}  The difference in energy  between two states of the Hydrogen atom that would vanish if the Schroedinger or Dirac equation are used.

\term{Lepton} the general term for a particle, such as the muon or electron that does not undergo the strong interaction.  
 
 \term{Lepton Universality} a principle of the standard model  stating that the only differences between observations of the effects of leptons are due to the different masses.
 
 \term{Standard Model} the current paradigm that  describes  quarks, leptons, vector bosons  and the electromagnetic, weak, and strong interactions between them.
\end{glossary}

\begin{glossary}[Nomenclature]
\begin{tabular}{@{}lp{34pc}@{}}
QCD& Quantum Chromodynamics\\
QED& Quantum Electrodynamics\\
CODATA & Committee on Data for Science and Technology \\
MAMI & Mainz Microtron\\
$\sigma^2$ &  mean square deviation of a distribution from its mean, the uncertainty of a measurement\\
 \end{tabular}
\end{glossary}

\begin{abstract}[Abstract]
 Pohl et al.  measured the energy difference between the 2P and 2S states of muonic
hydrogen and used it  to determine a precise value of the proton radius. The result disagreed significantly
from values  extracted from electronic hydrogen and  elastic electron-proton scattering. This  discrepancy was exciting because it indicated a breakdown of Coulomb's law. In more technical terms, the discrepancy indicated that a fundamental property of the Standard Model, known as lepton universality, could be  violated. This chapter explains the origins, meaning and significance of the puzzle. A resolution, based on very recent experiments,  is stated.   The proton radius puzzle is no more.
\end{abstract}

\begin{BoxTypeA}[chap1:box1]

\underline{Key points}

\begin{itemize}
\item The charge radius of the proton can be determined using electron- or muon-proton scattering at very low momentum transfer or  very accurate measurements of the spectra of the hydrogen atom.

\item The precision of atomic spectroscopy was improved in 2010 and 2013 by using muonic hydrogen. The value of the proton radius so obtained was significantly smaller than previous measurements. This puzzle engendered a huge response from the physics community.

\item Recent results from improved electronic hydrogen spectroscopy  measurements obtain values of the radius in agreement with that obtained from muonic atoms, with the very latest  measurement done at very high precision.
There is no longer a discrepancy between radii obtained using muonic or electronic hydrogen spectroscopy. The radius of the proton is the small value of about 0.84 fm.

\item Electron and muon scattering experiments are currently in progress, with the aim of improving their accuracy and testing lepton universality.

\end{itemize}
\end{BoxTypeA}
\section{Introduction}\label{chap1:sec1}

 The  determination~(\cite{Pohl:2010:Nature_mup1})  of the proton radius from  the measured  Lamb shift (energy difference between the 2P and 2S states) in the muonic hydrogen atom astounded the world of physics. The value of 0.84184(67)   fm differed by  about 4 \%  or 4.4 standard deviations from the electronic hydrogen atom  CODATA (Committee on Data for Science and Technology)~(\cite{Mohr:2012:CODATA10}) value of 0.8758(77) fm and from a
similar value with larger uncertainties determined by electron scattering~(\cite{Bernauer:2010:NewMainz}).  A later more precise muonic measurement~(\cite{Antognini:2013:Science_mup2}) found a 7.2 standard deviation difference. These differences became known as the proton radius puzzle~(\cite{Miller:2011yw}). This puzzle  brought up several interesting questions. The most prominent are:  what is the proton radius, why or is the electron-proton Coulomb interaction different than the muon-proton one, how can atomic physics (with relevant distance scales of Angstroms be used to measure an important  property of the proton (with relevant distance scales femtometers), why should muonic hydrogen be more sensitive to this quantity than electronic hydrogen, is  a 4 \%  difference really big, and finally was it really a puzzle? 

I use the  term `small radius' for the results of the muon-hydrogen experiment and the term  `large radius' for  other results prior to the discovery of the small radius. The proton radius puzzle is therefore the question: is the proton radius small or large?

\subsection{What is the proton radius?}\label{chap1:subsec1}

The proton, with its constituents of quarks, anti-quarks and gluons is a complicated, very small, quantum mechanical object, with size of  one part in one hundred thousand of that of a typical atom. If an atom was the size of a football field, the proton would be about  the  size of a pea. 
Quantum mechanics teaches us that any interaction used to probe  a system that small disturbs  the system. This means that one cannot simply 
use a ruler to measure the size of a  proton. So the question, ``what is the proton radius?" brings to mind the quotation,
``beauty is in the eye of the beholder".  The value of the radius depends on the probe used to measure the property. Here, the probe is the electric charge interaction observed at low momentum transfer. Other probes related to magnetic,  weak or gravitational interactions will  yield different values of the radius. To fully understand the proton it is necessary to understand the origins of the different radii, but the  radius
most accessible by several different  techniques is the charge radius.

I follow ~(\cite{Miller:2018ybm})  to show how the effect of the non-zero size of the proton impacts the energy levels of the hydrogen atom.  The very elaborate standard procedure  for handling corrections to the non-relativistic treatment of hydrogen spectroscopy with point-like protons is well-documented~(\cite{Eides:2000xc,Eides:2007exa}). The essential aspects are discussed here. The non-relativistic Hamiltonian for a system of a point-like proton and a lepton (muon or electron)  with a  
Coulomb interaction is given by 
\begin{align}\label{chap1:eq1}
H_0={\vec p^2\over 2m_r}-{\a\over r}
\end{align}
in the center of mass system, where $\a$ is the fine structure constant  ($\approx  1/137.035$) and $m_r$ is the lepton-proton reduced mass.
The size of the system is governed by the Bohr radius: $a_0=1/(\m_r \a)\approx 5.29 \times  10^{-11}$ m for electronic atoms
 and about 200 times smaller for muonic ones.
This starting Hamiltonian is improved by including a set of well-known correction terms. The first is that one solves the Dirac equation instead of the Schroedinger equation.  The most important effect of the proton's size is that the lepton-proton interaction contains a factor $G_E(-q^2)$ where $G_E$ is the Sachs  electric form factor~(\cite{Ernst:1960zza,Sachs:1962zzc})   and $-q^2=Q^2 =q_0^2-\vec q^2$ is the square of the momentum transferred in the lepton-proton  interaction. The typical value of the momentum transfer is the inverse size of the atom
$|\vec q|\approx 1/a_0$. The size of the energy transfer is $\vec q^2/(2M_p)$ so that $q^0/|\vec q| \approx 2\times 10^{-4}$, so that $-q^2=\vec q^2$ to within one part in $10^{-8}$ for electrons and $\sim 10^{-4}$ for muons, negligible in either case.  Since $|\vec q|$ is itself small one may 
 use the lowest-order Taylor-series  expansion to extract the effect of  the proton size  from  measured energy levels. Thus the replacement:
\begin{align}\label{slope} G_E(\vec q^2)=1 +\vec q^2 G_E'(0).
\end{align}
is made in atomic physics so that
  the Fourier-transform of the Coulomb interaction becomes:
\begin{align} V_C(\vec q^2)=-{4\pi\a\over \vec q^2}(1+\vec q^2G_E'(0))
= -{4\pi\a}({1\over  \vq^2}+G_E'(0)),
\label{vq}
\end{align} 
and  the coordinate space potential, $V_C(r),$ is    the three-dimensional Fourier transform of the previous expression with the result:
\begin{align}V_C(r)= -{\a\over r}-{4\pi\a}G_E'(0)\d(\vec r). \label{vr}
\end{align}  
Thus the Coulomb potential, usually thought of as being of long range, has a short-ranged, $\d(\vec r)$ component because of the non-zero proton size. Moreover, 
the non-zero proton size   causes  a repulsive correction to the Coulomb potential opposing  the usual electron-proton interaction. This is 
because  $G_E$ falls with increasing $\vq^2$ for small values of the momentum transfer. The appearance of the Dirac delta function causes the correction to occur only in S-states and generates a difference between S and P states, a contribution to the Lamb shift.
 
The typical value of $\vq^2$ is of the order of $1/a_0^2$.  The muonic hydrogen atom Bohr radius is about 200 times smaller than  that for the electronic one, but this huge difference does not appear in the modified Coulomb potential of \eq{vr}.

The shift in the energy, $\D E$,  is given by the matrix element:
\begin{align} \D E=\la \p_{nl} |\D V_C|\p_{nl}\ra=
 -{4\pi}\a G_E'(0) |\p_{n0}(0)|^2\d_{l0},\label{good}\end{align}
 where $\p_{n0}$ is the wave function of the $n$'th excited state.
The  effect of the non-zero size of the proton  on the  energy shift is determined by the slope of $G_E$  at its origin. 
 The effect, \eq{good}, is of order $\a^4$ because $ |\p_{n0}(0)|^2$ is of order $\a^3$, so that this term is  included with the others of order $\a^4$.
 
 For historical reasons, discussed in the next Section, the slope is defined as 
\begin{align}\label{def} G'_E(0)=-{r_p^2\over 6},\end{align}   so that
\begin{align}\D E={4\pi}\a {r_p^2\over 6} |\p_{n0}(0)|^2\d_{l0}\end{align} 
is  used to analyze the data. 

The probability density at the origin $|\p_{n0}(0)|^2$ is proportional to $1/a_0^3 $ and thus   the cube of the mass of the electron or muon. The muon is about two hundred times more massive than the electron, so the effect on the energy shift is 8 million times larger for the muon than for the electron.  This explains how  a moderately precise measurement in muonic hydrogen  can lead to a huge improvement in the accuracy of the proton radius determination.

\subsection{Why should a 4\% difference matter?}
Coulomb's Law, which is responsible for chemistry,  states that the force between two particles depends on the product of the charges and the distance between them.  
Finding a difference between the muon-proton and electron-proton force would violate electromagnetic physics in a major way. More technically, such a difference would violate the well-tested principle of lepton universality and be physics beyond the Standard Model.

\section{Electron Measurements Prior to the 2010 Muon Measurement}\label{chap1:sec2}

This section addresses the question of why the 2010 Muon Measurement was such a surprise. 

\subsection{Electron-proton scattering}

Elastic electron scattering has been used to measure the
electromagnetic structure of nucleons and nuclei since the pioneering work of Hoftstatder~(\cite{Hofstadter:1955ae}). Reviews  since that  time  include~(\cite{Perdrisat:2006hj,Arrington:2011kb}). The proton structure information is encoded in the Sachs electric, $G_E$ and magnetic $G_M$  electromagnetic form factors. There are two form factors because the proton has both charge and spin, with the latter causing magnetic properties.

\begin{figure}[h]
\begin{center}
\includegraphics[width = 0.4\columnwidth]{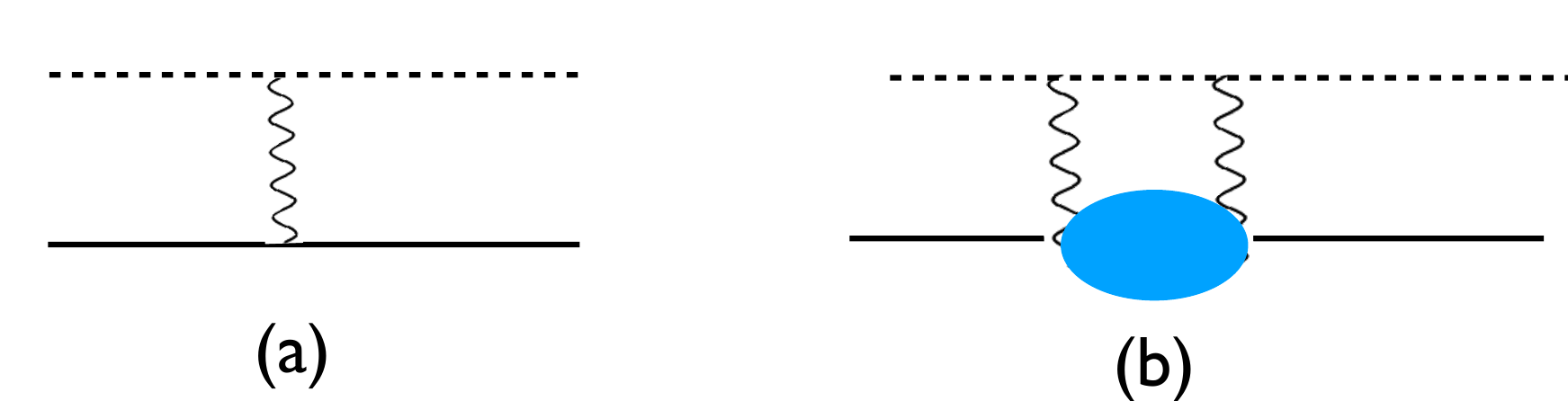}
\caption{ Lepton (dashed line)  scattering from a proton (solid line) via exchange of one or two photons (curved line). (a) One-photon exchange (b) Two photon exchange. The blob represents an intermediate excited state. The graph with the two photons interchanged and crossing  is not shown.  %
 }
\label{2g}
\end{center}
\end{figure}
The lepton-proton scattering process is depicted in Fig.~(\ref{2g}). The dominant one photon-exchange amplitude is obtained from knowledge of $G_{E,M}$ and well-known terms. The two photon-exchange term is more complicated because the proton is pushed into an excited state by the first photon and pushed back into the proton state by the second photon. The two-photon exchange term is generally smaller than the one photon exchange term by a factor of the fine structure constant $\a\approx 1/137$, but must be kept if achieving high accuracy is necessary. The importance of the  two-photon exchange process was not recognized until 2003~(\cite{Guichon:2003qm}). The two-photon exchange amplitude has the same sign for $l^\pm$-proton scattering ($l=e,\mu$), but the one-photon exchange  term changes sign when the charge changes sign.
Thus the  interference term between the one- and two-photon exchange amplitudes, proportional to their product  changes sign between negatively and positively charged lepton-proton scattering. Taking the average of $l^+-p$ and $l^--p$ scattering approximately removes the two-photon exchange effect.

The proton radius  is defined according to \eq{def}. This means that obtaining data at small values of the momentum transfer $Q^2 $  is very  important, causing a variety  of well-known issues~(\cite{Rosenfelder:1999cd,Sick:2003:RP}). 
Ideally, one would like to 
 use   the Taylor series expansion 
\begin{align}
G_E(Q^2) = 1 - Q^2 r_p^2 / 6 + Q^4r_p^4 / 120 +\ldots   
\label{Taylor}
\end{align}
to obtain a
model independent radius determination. The measurement of the term 1 at the necessary precision requires knowing the beam intensity and target properties to an extraordinarily high  accuracy. Obtaining a significant difference from unity requires
momentum transfers of the order of 1 fm$^{-2}$ which means that  $Q^2$ is large enough so that there has been no range of $Q^2$ where the $r_p^2$ term dominates over the 1 or the $r_p^4 $ term. 
The above expression is not used in practice. Instead, one uses models of various functional forms, and one must be concerned with an uncertainty due to  model dependence.
For example, different extractions of $r_p$ from
 the data that produced the large radius ~(\cite{Bernauer:2010:NewMainz}) may  yield the small radius, for example, ~(\cite{Alarcon:2020kcz}).  
The hydrogen atom physics accesses momentum transfers of order $1/a_0^2= 3.5 \times 10^{-10} \,\rm fm^{-2} $ for 
electrons and  $1.4 \times 10^{-5}  \rm fm^{-2}$ for muons. These values are much smaller than those available in typical electron scattering experiments,  a complete mismatch of scales between the scattering and hydrogen spectroscopy methods to obtain $r_p$.
This means that experiments based on hydrogen spectroscopy typically  have greater leverage in obtaining precision values of $r_p$.
  Therefore, the values of $r_p^2$ obtained from electron scattering  seem to depend significantly on the  extrapolation techniques employed. Recent attempts to improve the situation are discussed below.

\begin{figure}[h]
\begin{center}
\includegraphics[width = 0.6\columnwidth]{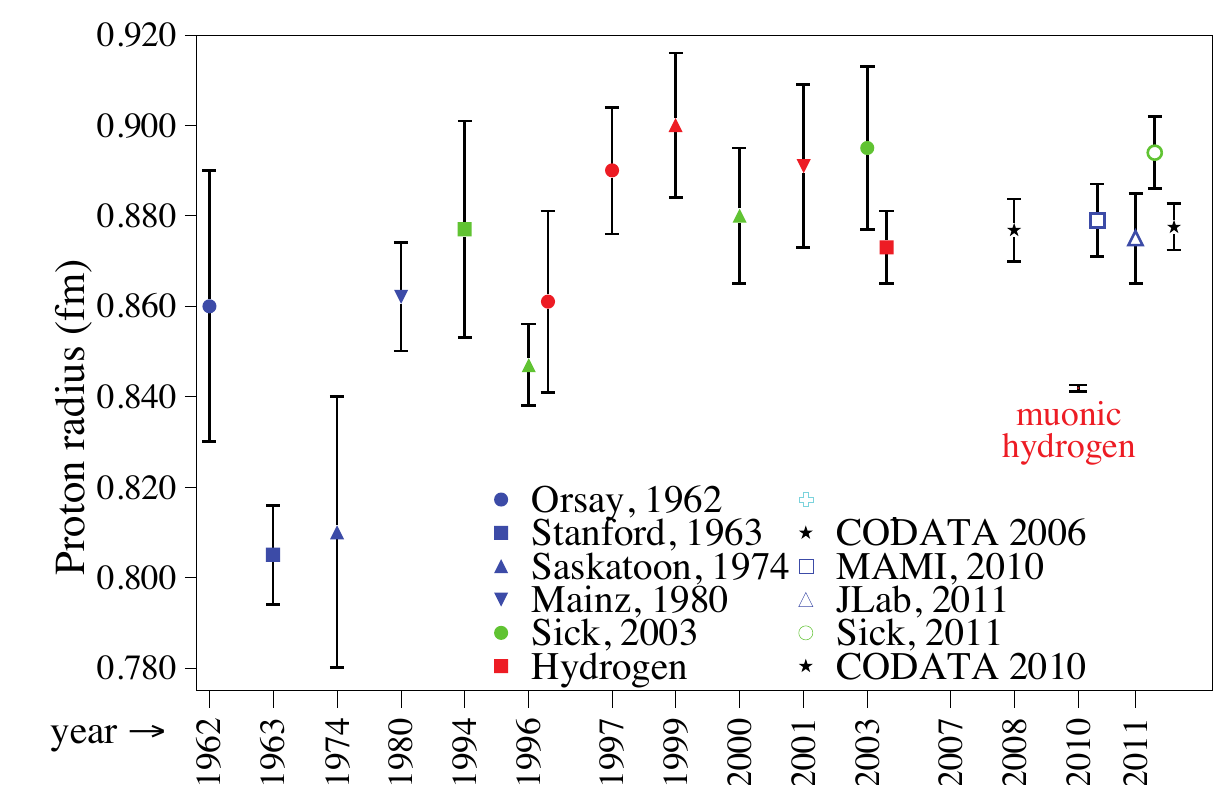}
\caption{The  proton radius as of 2013~(\cite{Pohl:2013yb}). 
  The values obtained  from early electron scattering measurements  are:
  Orsay~(\cite{Lehmann:1962:RP_Orsay}),
  Stanford~(\cite{Hand:1963:RP_Stanford}),
  Saskatoon~(\cite{Murphy:1974:RP_Saskatoon,Murphy:1974:RP_Saskatoon_erratum}),
  Mainz~(\cite{Simon:1980:RP_Mainz}) (all in blue) .
Newer  scattering measurements are from 
  MAMI~\cite{Bernauer:2010:NewMainz} and 
  Jlab~\cite{Zhan:2011:JLab_Rp}.
  The green and cyan points denote various reanalyses of the world electron 
  scattering data~(\cite{Wong:1994sy,Mergell:1995bf,Rosenfelder:1999cd,
    Sick:2003:RP,Belushkin:2007:NuclFF,Sick:2011:Troubles}).
  The red symbols display values of \rp{} obtained from laser spectroscopy of 
  atomic hydrogen and advances in hydrogen QED theory (see 
 ( \cite{Mohr:2012:CODATA10}) and references therein).
  The green and red points in the year 2003 denote the reanalysis of the world 
  electron scattering data~(\cite{Sick:2003:RP}) and the world data from hydrogen
  and deuterium spectroscopy which have determined the value of \rp{} in the 
  CODATA adjustments~(\cite{Mohr:2008:CODAT06,Mohr:2012:CODATA10})
  since the 2002 edition.
  %
}
\label{fig1:Rp_vs_t}
\end{center}
\end{figure}

Fig.~\ref{fig1:Rp_vs_t} shows several determinations of the proton charge 
radius available at the time of the  announcement of the muonic atom results.
 Early elastic electron-proton scattering measurements from
Orsay~(\cite{Lehmann:1962:RP_Orsay}), Stanford~(\cite{Hand:1963:RP_Stanford}),
Saskatoon~(\cite{Murphy:1974:RP_Saskatoon,Murphy:1974:RP_Saskatoon_erratum}) and
Mainz~(\cite{Simon:1980:RP_Mainz}), and the various re-analyses of these world
data~(\cite{Sick:2003:RP,Blunden:2005:RP}) are displayed.  Spectroscopy of atomic hydrogen
became sensitive to the value of \rp{} at the percent level in the mid-1990. The  value from CODATA (The Committee on Data for Science and Technology)  that  recommends 
basic constants and conversion factors of physics and chemistry for international use (\cite{Mohr:2012:CODATA10}) is also shown.
 Electronic measurements
gave values of about  \rp=0.88\,fm, but  the value obtained from muonic hydrogen
  (\cite{Pohl:2010:Nature_mup1,Antognini:2013:Science_mup2})  is \rp=0.84\,fm. This seemingly small difference is actually startling when expressed in terms of the frequency of the detected light emitted from a muonic atom, see Fig.~\ref{fig3:mup_resonance} below.
  
\subsection{Hydrogen Spectroscopy}
There is a long and amazing history of unanticipated findings.
Discovering  that the emission of light from decaying atoms was quantized led to the development of quantum mechanics. 
The discovery of
the Lamb shift in hydrogen~(\cite{Lamb:1947:FShyd})  found the first effects of
quantum electrodynamics   that  remove the degeneracy (equality) of energy levels predicted by the Dirac and Schroedinger equations.

Quantum electrodynamics (QED) describes the energy levels of hydrogen  with extraordinary accuracy. The test of
QED using measured transition frequencies in hydrogen is limited by two of the necessary input
parameters. The first is   the Rydberg constant
\Ryd{}. This enters in the energy difference between states of different principal quantum number, $n$.
The 1S-2S transition in hydrogen has been measured with an accuracy of 4 parts
in $10^{15}$~(\cite{Parthey:2011:PRL_H1S2S}).   Their value has  been used to determine the Rydberg constant, \Ryd{}.
 See~(\cite{Mohr:2012:CODATA10}) for details 
The second input parameter  is  the   proton charge radius \rp{}.
Once the values of \Ryd{} and \rp{} are known QED bound-state calculations can be tested to high accuracy.

The energies of S-states, in simplified form, are given by
\begin{equation}
\label{eq:E_simple}
E(nS) \simeq - \frac{\Ryd}{n^2} + \frac{L_{1S}}{n^3},
\end{equation}
where $n$ is the principal quantum number, and $L_{1S}$ denotes the Lamb shift
of the 1S ground state caused  by QED as well as  the effect of the
proton charge radius \rp{}. Both of these effects are proportional to $ |\p_{n0}(0)|^2$ that is itself proportional to $1/n^3$.
In muonic hydrogen 
\begin{align} L_{1S} \simeq ( 8172 + 1.56\,\rp^2 )\,\rm MHz,
\label{Lamb}
\end{align} 
when \rp{} is expressed in fm  so the finite size effect on the 1S
level in hydrogen  is about 0.15\,MHz.
The two terms of Eq.~(\ref{eq:E_simple}) depend differently on $n$. Thus, if at least 2  transition frequencies are measured in hydogen, one  can solve the  high school math problem of two equations and  two unknowns to  determine the values of both \Ryd and \rp.

\begin{figure}[t]
\begin{center}
\includegraphics[width = .60\columnwidth]{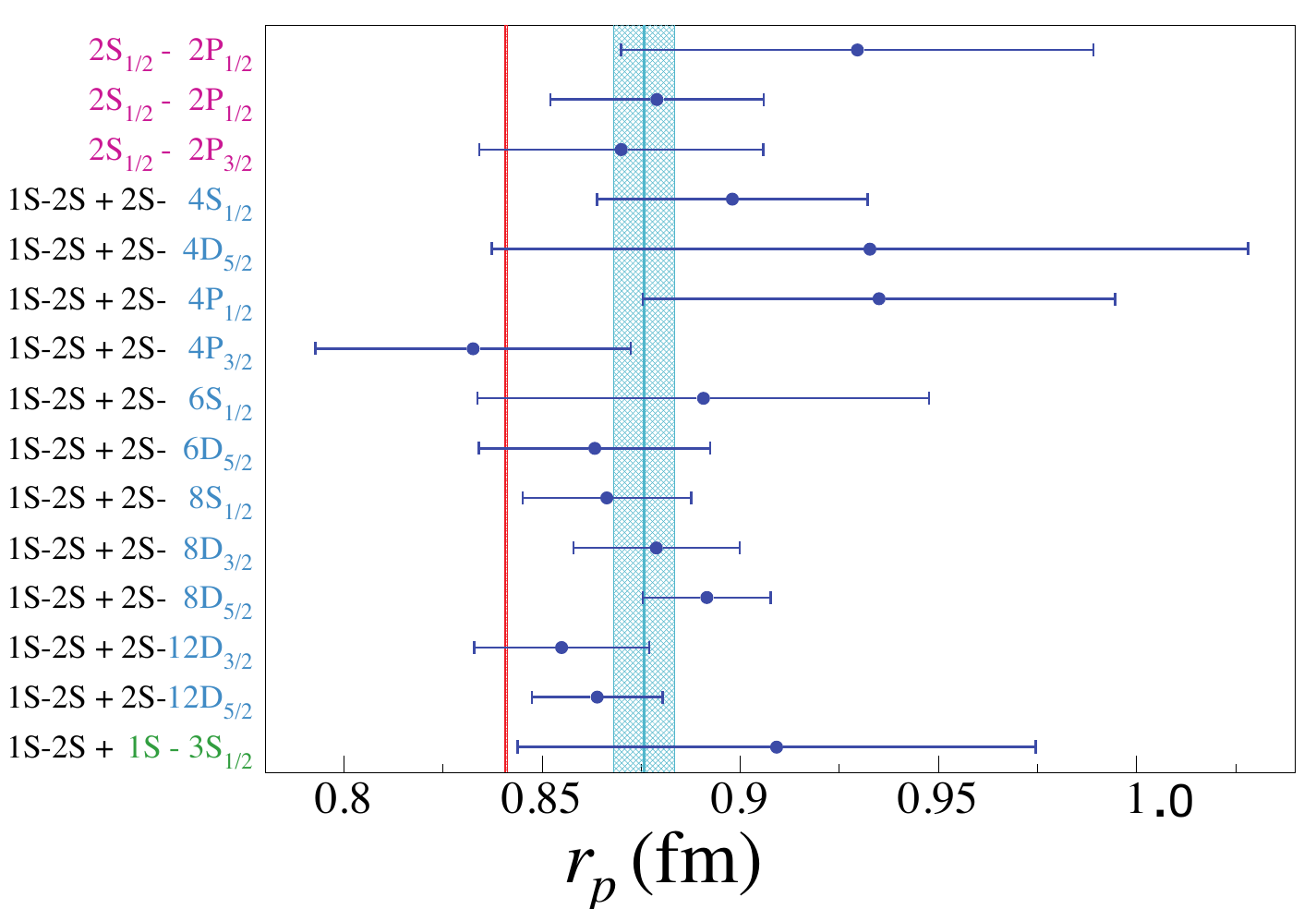}
\caption{Proton charge radii \rp{} obtained from hydrogen
  spectroscopy~(\cite{Pohl:2013yb}). 
  The value from muonic
  hydrogen~(\cite{Pohl:2010:Nature_mup1,Antognini:2013:Science_mup2}) ($r_p=0.84087\pm 0.00039$ fm) is shown with its error bar as the vertical red line. The average of the electronic results is $r_p=0.8758\pm 0.0077\,  fm$ and is shown in the blue rectangle.}
\label{fig2:Rp_from_H}
\end{center}
\end{figure}

It was typical to use  the most accurately measured 1S-2S
transition~(\cite{Parthey:2011:PRL_H1S2S}) and one of the 2S-8S,D/12D
transitions~(\cite{deBeauvoir:1997zz,Schwob:1999:Hydr2S12D}).  
The former contains the largest 1S Lamb shift $L_{1S}$ and so is most
sensitive to the value of \rp{}. The latter contains only smaller Lamb shift contributions
due to the $1/n^3$ scaling in Eq.~(\ref{eq:E_simple}), and hence determines
\Ryd{}. Fig.~\ref{fig2:Rp_from_H} shows the different values of \rp{} obtained
by combining the 1S-2S transition and each of the other precisely measured
transitions in hydrogen. In addition, it contains three values of \rp{} obtained from
a direct measurement of the 2S-2P transitions in hydrogen that do not depend on
the Rydberg constant. The consistent use of the 1S-2S transition causes the different measurements to lack statistical independence. The tiny uncertainty of 0.0077 fm is obtained by assuming statistical independence of 15 measurements and so is   likely an underestimate. Using only the three $2S_{1/2}-2P_{1/2}$ measurements leads to a $r_p=0.868\pm0.022$ which is nearly consistent within one standard deviation with the muonic hydrogen value of $r_p=0.84087\pm 0.00039$ fm. Even in 2012, a closer look at the data could have made the puzzle seem smaller.
See also (\cite{Mergell:1995bf}).

\section{The Muonic Hydrogen Measurement}
The  Charge Radius Experiment with Muonic Atoms (CREMA) collaboration carried out  the experiment. 
 Learning the  Lamb shift requires measuring the energy differences of the 2S-2P transitions.  Laser spectroscopy is the tool of choice.  The first challenge here is  to make      \mup{}
atoms in the metastable 2S state. The first observation of long-lived \mup{} atoms in the 2S
state~(\cite{Kottmann:1999:EXATconfProc,Pohl:2001:MolecQuenchMup,Pohl:2006:MupLL2S})
ultimately led to  
the successful Lamb shift measurement~(\cite{Pohl:2010:Nature_mup1,Antognini:2013:Science_mup2}).

A new beam line for low-energy $\mu^-$~(\cite{Kottmann:2001:Conf:Trieste}), delivering an order of magnitude more muons than previously, was constructed at the Paul-Scherrer-Institute
(PSI) in Switzerland.   About half of the muons stop in
the  target vessel.   Two stacks of ultra-thin carbon foils  are used to  detect the muon with high efficiency. A signal in both stacks starts the data acquisition and the
pulsed laser system.

\begin{figure}[h]
\bigskip
\begin{center}
\includegraphics[width = .60\columnwidth]{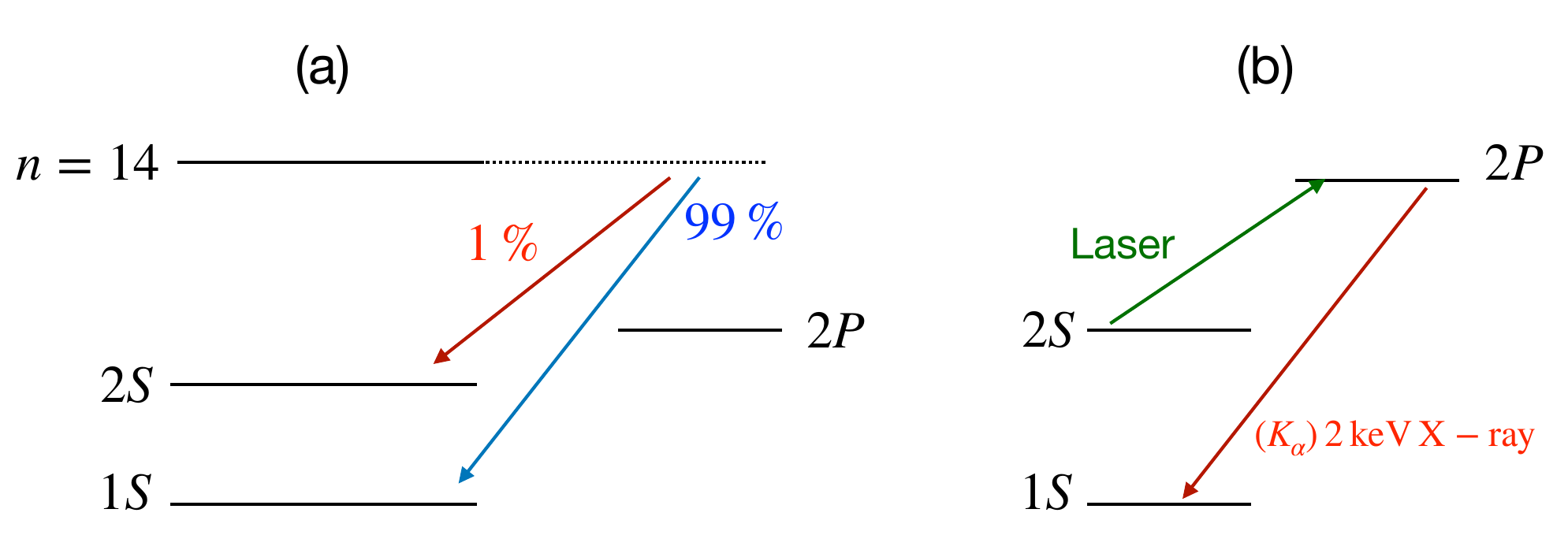}
\caption{Energy levels  and experimental principle in muonic hydrogen. (a) 1\%  of the muons are captured in the metastable 2S state, with the rest  captured in the 1S ground state, emitting prompt X-rays (blue). (b) The $\mup(2S)$ atoms are illuminated by a laser pulse (green). If the laser frequency matches that of the Lamb shift  X-rays are observed (red)}
\label{fig3:mup_spectra}
\end{center}
\end{figure}

When negatively charged muons are stopped in molecular \Htwo{} gas at low pressure
about 1\% of the muons form \mup(2S)
atoms~(\cite{Kottmann:1999:EXATconfProc,Pohl:2009:EXA08}) of  lifetime of
about 1\,\mus~(\cite{Pohl:2006:MupLL2S}), see  Fig.~\ref{fig3:mup_spectra}.
A pulsed laser is used to excite the atom from the 2S state to the 2P state.
The pulsed laser
system~(\cite{Antognini:2005:6mumLaser,Antognini:2009:Disklaser}) is  used to amplify
red light. These  pulses
are converted to the desired infrared wavelength at 5.5-6\,\mum{}. %
The frequency of the laser light is   
calibrated in the infrared region using well-known absorption lines of water vapor
(\Water{}). The
final uncertainty of the laser frequency
calibration is 300-MHz~(\cite{Pohl:2010:Nature_mup1}).

The successful 2S-2P laser excitation is signaled by the emission of a
1.9\,keV \Ka{} x-ray (transition from $n=2$ to $n=1$),  emitted in the radiative 2P-1S de-excitation that
immediately follows the 2S-2P transition. The x-rays are detected in   photo diodes ~(\cite{Fernandes:2003:NIM,Ludhova:2005:LAAPDs}).

\begin{figure}[h]
\bigskip
\begin{center}
\includegraphics[width = .80\columnwidth]{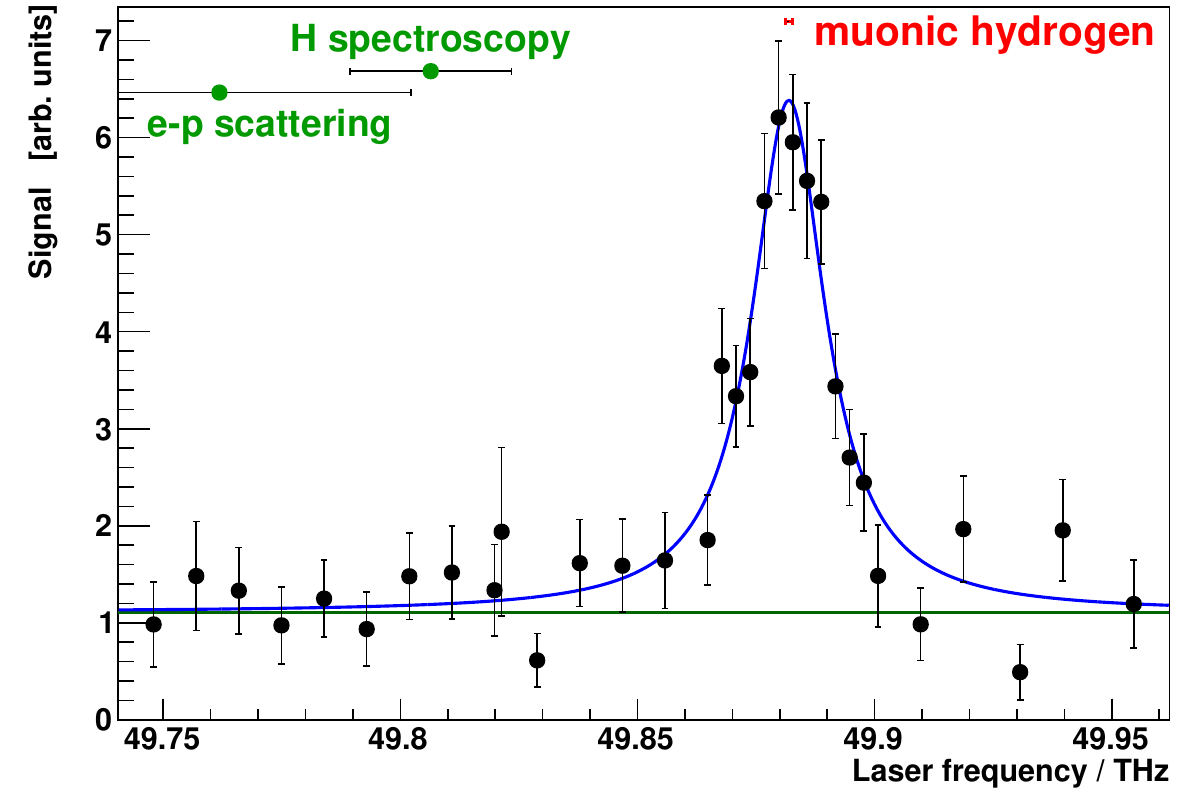}
\caption{Resonance in muonic hydrogen, together with the positions predicted 
using the proton radii from elastic electron-proton scattering using pre-2009
world data~(\cite{Sick:2003:RP,Blunden:2005:RP}) and the CODATA-2006 value from
H spectroscopy~(\cite{Mohr:2008:CODAT06}).
}
\label{fig3:mup_resonance}
\end{center}
\end{figure}

The (blue) resonance curve in Fig.~\ref{fig3:mup_resonance} 
is obtained by plotting the normalized number of
laser-induced \Ka{} x-rays as a function of laser frequency. The resonance frequency and  the  resulting Lamb shift 
is revealed by a fit of this
 curve~(\cite{Pohl:2010:Nature_mup1,Antognini:2013:Science_mup2}).

The difference between the observed frequency and the expected frequency was startling and surprising. The initial run of the experiment had been tuned to frequencies corresponding to the larger radius and no signal was found. The experimental equipment had to be quickly readjusted to find the correct transition frequency.

The measured radius differed by 4.9 standard deviations from the CODATA's value. Later work suggested a 7.2 standard deviation difference~(\cite{Antognini:2013:Science_mup2}).  This strange difference between values  of $r_p$ obtained by using electrons and muons, both  supposedly highly accurate evaluations, became known as the proton radius puzzle.

\section{Immediately Proposed  Solutions}
One possibility was that the initial muonic hydrogen experiment was in error. This was easily discounted; see ~\cite{Karr:2012:3body,Pohl:2013yb}.

The difference between the muonic and electronic determinations of the value of \rp{}  prompted a huge reaction of theorists from   different fields with an interesting sociology. Some QED physicists proposed a solution involving QED, the nuclear physicists proposed ideas from strong interaction physics and the particle physicists proposed  the existence of new particles.  

The relevant QED theory was carefully reviewed in~(\cite{Pohl:2013yb}) and~(\cite{Antognini:2013rsa}) and there are no  effects that change the conclusions.

The nuclear physics effects involved the two-photon exchange contribution to the lepton-proton interaction~(\cite 
{Miller:2011yw,Miller:2012ne}). These effects vary as the cube of the nuclear charge, and were ruled out by observations of muonic helium ion ~(\cite{Krauth:2021foz}) as well as electron-nucleus quasi-elastic scattering~(\cite{Miller:2012:quasi-elastic}).

The most interesting possibility is that the difference in the radius measurements is related to a fundamental difference between the $e-p$ and $\mu-p$ interaction. According to the Standard Model the only differences are those that originate in the mass difference. Any other difference between the two interactions would be  a violation of lepton universality, a basic principle of the Standard Model. At the same time there was another muonic puzzle: an almost   3$\sigma$ difference between the measurement of the anomalous magnetic moment~(\cite{Muong}) and the Standard Model prediction
~(\cite{Jegerlehner:2009ry}). It was tempting for theorists to try to explain both puzzles using  the same new particles ~(\cite{Barger:2011,Tucker-Smith:2010wdq,Batell:2011qq,Liu:2016qwd,Liu:2016mqv}). No new particles have been discovered. Furthermore,   there is now agreement  between the measurement and Standard Model theory values of the muon anomalous magnetic moment~(\cite{Hertzog:2025ssc,Aliberti:2025beg}). This finding greatly reduces the sizes of possible effects of any postulates  involving new particles.

\section{Experimenters Responses to  the 2010 Muon Experiment}

The proton size puzzle was broadly  reviewed in 2020~(\cite{Karr:2020})  and 2022~(\cite{Gao:2021sml}) where detailed discussions of  both hydrogen spectroscopy and scattering are  presented.  An update is provided here.
The results of spectroscopy are discussed first, followed by descriptions of published and planned lepton-proton scattering experiments.
\subsection{Hydrogen Spectroscopy}

The existence of the puzzle stimulated much elegant efforts in precision electronic hydrogen spectroscopy,
~(\cite{Beyer:2017,Fleurbaey:2018fih,Bezginov:2019mdi,Grinin:2020txk,Brandt:2021yor,Maisenbacher:2026nau}).
A summary of these experiments is shown in Table 1.

\begin{table}[t]
\TBL{\caption{Electron hydrogen experiments since 2016: The average is $r_p=0.843\pm 0.0013\,$ fm.}\label{chap1:tab1}}
{\begin{tabular*}{\textwidth}{@{\extracolsep{\fill}}@{}lll@{}}
\toprule
\multicolumn{1}{@{}l}{\TCH{Reference}} &
\multicolumn{1}{c}{\TCH{Transition}} &
\multicolumn{1}{l}{\TCH{$r_p$ fm}}\\
\colrule
\cite{Beyer:2017} & 2S-4P  & 0.8335 (95)\\
\cite{Fleurbaey:2018fih} & 1S-3S & 0.877(13)\\
\cite{Bezginov:2019mdi} &2S-2P& 0.833(10)\\
\cite{Grinin:2020txk} & 1S-3S &  0.8482(38)\\
\cite{Brandt:2021yor}&$\rm 2S_{1/2}-8D_{5/2}$&0.8584 (51)\\
\cite{Maisenbacher:2026nau}& 2S-6P&0.8406 (15)\\
\cite{Bullis:2026ebw}&2S-nS&0.8433(31)\\
\botrule
\end{tabular*}}{%
}%
\label{table}
\end{table}
 
 The average of these results is $r_p=0.843\pm 0.0013\,$ fm, dominated by the most recent measurement~(\cite{Maisenbacher:2026nau}). This measurement, ``at least 2.5-fold more precise than from other atomic hydrogen determinations",  finds a ``Sub-part-per-trillion test of the Standard Model with atomic hydrogen" which is the most precise test of bound-state QED calculations. The value of the proton radius no longer impacts tests of QED.
 
\begin{figure}[h]
\bigskip
\begin{center}
\includegraphics[width = .80\columnwidth]{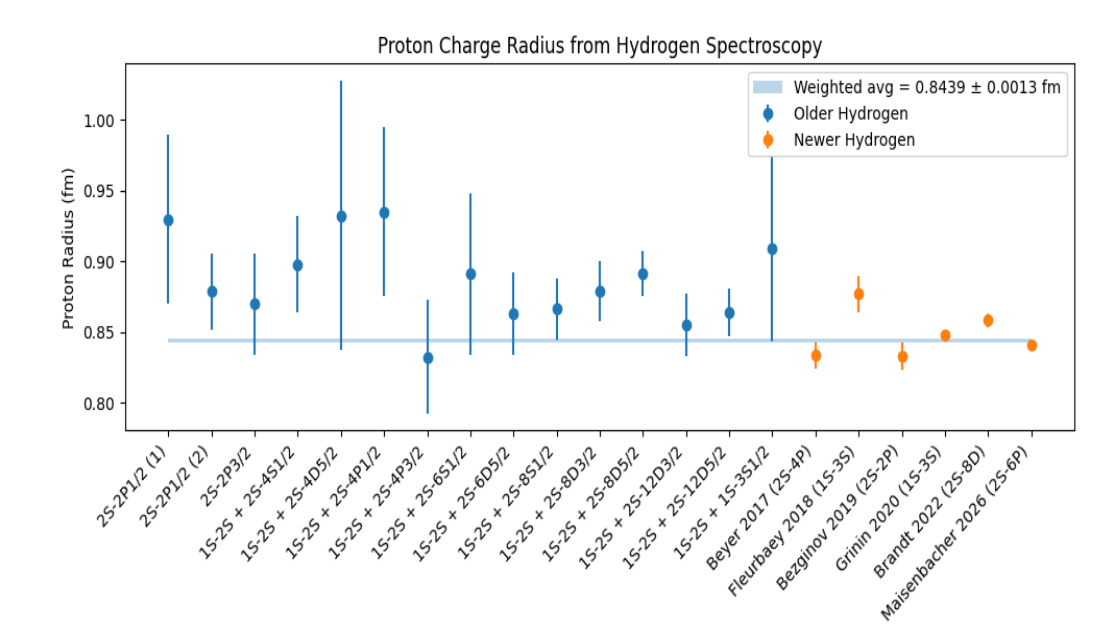}
\caption{Proton  charge radius determinations from electronic  hydrogen spectroscopy. Blue points denote older transition measurements, while orange points show recent  results (2017-2026). Error bars represent quoted 1$\sigma$ uncertainties. The shaded band indicates the inverse-variance weighted average over all measurements, yielding
$r_p=0.8439\pm
0.0013$ fm.
}
\label{final}
\end{center}
\end{figure}

The data shown in Table~1 are much more precise than shown in Fig.~3 and dominate a combined average.  The effects of correlations between different measurements seem  to be small in tests done by the author and E.~Slate.
 See Fig.~\ref{final}.
There is now no difference between muonic and electronic hydrogen determinations of  $r_p$.

\subsection{Lepton-proton Scattering experiments}

  I start by describing the results of published experiments.
\subsubsection{Published electron-proton scattering experiments}

The  results from MAMI (Mainz Microtron)~(\cite{Bernauer:2010:NewMainz}) were updated and a  new determination of the electric and magnetic form factors made that included the effects of two-photon exchange~(\cite{A1:2013fsc}). This also included all of the previous world data. There was little change to the previously determined value of $r_p$. 
Further work at Mainz~(\cite{A1:2022wzx})  measured electron-proton scattering at small values of $Q^2$ between 0.25 fm$^{-2}$ and 1.1 fm$^{-2}$
 using a newly developed cryogenic supersonic gas jet target. The results agreed with the previous  measurements made at Mainz.  
 A different novel   initial state radiation (ISR) technique was also used at Mainz~(\cite{Mihovilovic:2016rkr}) to 
 make a measurement of the proton's charge form factor at extremely low momentum transfer values.
The ISR method exploits the radiative tail of elastic electron-proton scattering peaks to  allow measurements $0.001\le Q^2\le \rm fm^{-2}$. 
This work obtained a proton radius  value of $r_p\approx 0.810 \pm 0.074 $ fm, with the uncertainty primarily arising from statistical uncertainties. This  accuracy  is not sufficient to distinguish between  known  differences in extractions.
 
The proton charge radius experiment (PRad) in Hall B at Jefferson Lab~(\cite{Xiong:2019umf})  used a unique 
magnetic-spectrometer-free calorimeter based method (allowing measurements at very small scattering angles)
  which along    with a
novel windowless hydrogen-gas-flow target allowed the experiment
to measure the e-p elastic scattering cross section in the 
unprecedentedly low  range: $5 \times 10^{-3}\le Q^2\le 1.25\,\rm fm^{-2}$.  The extracted  value of the charge radius
 is $r_p=0.831\pm 0.007 \,\rm (stat)\pm 0.012\,(syst)$ fm,  in agreement with the muonic hydrogen
spectroscopic results, and  also  the recent measurements from electronic  hydrogen Lamb shift displayed in Table 1.

\subsubsection{Planned Lepton-Proton Scattering Experiments}
The exciting need to validate or invalidate the small radius spawned many new ideas and technologies~(\cite{Gao:2021sml,Xiong:2023zih}). These experiments are listed in Table~\ref{lp}. A huge span of beam energies ranging between 10 MeV and 100 GeV will be employed at tiny values of momentum transfer to extract the proton radius.

\begin{figure}[h]
\bigskip
\begin{center}
\includegraphics[width = .9\columnwidth]{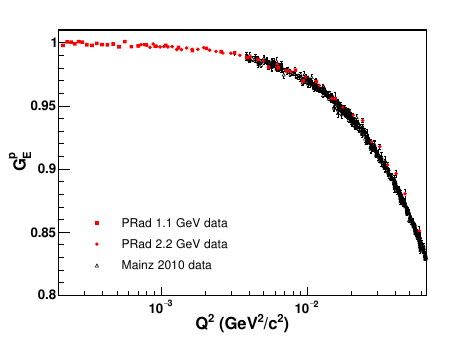}
\caption{ Comparison of PRad and Mainz extractions of the electric form factor, $G_E(Q^2)$.   From~(\cite{Xiong:2020kds,Gao:2024kkg}). The unit $\rm GeV^2/c^2$ is approximately equal to 25 fm$^{-2}$.
}
\label{comp}
\end{center}
\end{figure}

 The PRad result, within its experimental uncertainties, is in agreement with the small radius measured in muonic hydrogen spectroscopy experiments. Thus   the PRad result  conflicts with the modern electron scattering experiments. This result is 5.8\%, or three standard deviations, smaller than the value obtained in the most precise electron scattering experiment to date,~\cite{Bernauer:2010:NewMainz}.
A comparison of the PRad measurement with that of the Mainz result is shown in Fig.~(\ref{comp}).
The non-vanishing differences between the PRad and Mainz results at the larger values of $Q^2$ along with the realization that 
  PRad did not reach  the highest precision allowed by the calorimetric technique led to a new, upgraded experiment named
  PRad-II. This effort is  intended to reduce the overall experimental uncertainties by a factor of 3.8 compared to PRad and  will be the first lepton scattering experiment to reach the $Q^2$ range of $10^{-3} \rm fm^{-2}$  allowing a more accurate  extraction of $r_p$.  The PRad-II experiment with its projected total uncertainty of 0.43\% could demonstrate whether there is any systematic difference between $e-p$  scattering and muonic hydrogen results. This experimental run started March 26, 2026 and is currently scheduled until July 5, 2026.

The Muon Scattering Experiment (MUSE)~(\cite{MUSE:2013uhu,10.21468/SciPostPhysProc.5.023}) was stimulated by the exciting possibility that $ep$ scattering is fundamentally different than  $\mu p$ scattering. One should make a direct comparison of the scattering to see  if this is indeed the case. The MUSE  experiment  measures electron, muon and pion scattering at the same time.    Both positively and negatively  charged beams  of leptons will be used,  allowing the effects of two-photon exchange corrections to be determined. This experiment finished production data taking in Dec. 2025, with some systematic studies. continuing. 
Heavy baryon chiral perturbation theory calculations from~(\cite{Goswami:2025zoe}) suggest that the TPE can be as large as a several percent effect in MUSE kinematics, and larger for electrons than for muons.. 

The AMBER (Apparatus for Meson and Baryon Experimental Research) experiment (\cite{Ketzer:2026ytc}) intends to    provide a new, independent precision measurement
of $r_p$  by using high-energy muon-proton elastic scattering. A  high-energy (100 GeV)  muon beam
is available exclusively at the  CERN SPS.   The use of high energy has advantages including: the reduction  of several experimental systematic effects and smaller  radiative corrections. This experiment is the only one that measures the recoiling proton.  The two-photon exchange effects could be larger  than for the MUSE experiment due to the probability of exciting more intermediate states. The measurement
is ongoing.

The  MAGIX experiment (\cite{A1:2021njh}) at MAMI uses a unique   a windowless target design. A cryogenic supersonic gas jet target provides a relatively dense gas flow perpendicular to the electron beam direction, allowing  high luminosities.

A new measurement, the ULQ$^2$ experiment (\cite{Legris:2025lxo,Suda:2025cdl} of the proton and deuteron charge radii with low
energy electron scattering is being conducted in the Research Center for Accelerator
and Radioisotope Science (RARiS), Tohoku University, Japan. The 60 MeV
accelerator that provides  energies between 10 and 60 MeV will be used. The measurement can
be carried at scattering angles between $30^\circ$  and $150^\circ$ so that  the accessible regime of $Q^2$ can be as low as $7.5\times 10^{-7}\,\rm fm^{-2}$. The measurement of the electron-proton cross-sections were 
 completed during 2023 and 2024 in RARiS. The analysis is currently ongoing.

\begin{table}[t]
\TBL{\caption{Planned lepton-proton scattering experiments}\label{lp}}
{\begin{tabular*}{\textwidth}{@{\extracolsep{\fill}}@{}lll@{}}
\toprule
\multicolumn{1}{@{}l}{\TCH{Experiment}} &
\multicolumn{1}{c}{\TCH{Beam}} &
\multicolumn{1}{l}{\TCH{$Q^2\, \rm (fm^{-2})$}}\\
\colrule
PRad-II\,(\cite{PRad:2020oor})& $e^-$ & $10^{-3}-1.5$ \\
MUSE\,(\cite{Strauch:2018ros,Lin:2024gsb})& $e^\pm,\m^\pm$& $0.0375-2$\\
AMBER\, (\cite{Ketzer:2026ytc}) &$\mu^\pm$& 0.25-1\\
MAGIX\,(\cite{A1:2021njh}) & $e^-$ &  2.5$\times 10^{-3}-2.1$\\
ULQ$^2$\,(\cite{Legris:2025lxo,Suda:2025cdl})& $e^-$& 7.5$\times 10^{-3}-0.2$\\
\botrule
\end{tabular*}}{%
}%
\end{table}
 
It is useful to compare the equations (\ref{Taylor}) and (\ref{Lamb}). The   $\vec q^2r_p^2/6$ term in the former is of order
$10^{-4}$ compared to unity for the lowest planned value in the PRad-II experiment. In Eq.~(\ref{Lamb}) the   $1.56\,
\rp^2 $ compared to the leading term is also of the same ratio with respect to $L_{1S}$.
\section{Conclusions and Resolution of the Puzzle}

I begin by discussing  the status of the proton radius puzzle according to 
latest CODATA adjustment (dated  2022)~(\cite{Mohr:2024kco}).
They report that $r_p=0.84075 (64) $ fm, in agreement with the small radius obtained from muonic hydrogen,   if the complete data set is included. If no muonic atom data is included the value is $r_p=0.8529 (43) $ fm. They report the difference between the two determinations as being 2.8 $\sigma$ and that, 
``The proton radius 'puzzle' is not yet solved.". I believe that the very recent experiment~(\cite{Maisenbacher:2026nau}), with its  extraordinary precision and other recent measurements  have changed the situation. As shown in Table 1 and Fig.~3 there is now no disagreement between 
the electronic hydrogen and muonic hydrogen. Moreover, the most recent first-principles, lattice QCD calculations~(\cite{PhysRevD.109.094510}  find that the radius is small. 
The proton radius puzzle has been  solved by using improved measurements in hydrogen. The radius is small.

Despite this finding there is still much work to do.

\subsection{Electron-proton scattering}

The CODATA 
authors deemed that the  values of $r_p$ obtained from electron scattering
should not be included in the latest 
adjustment.   Their reasons involved  a lack of agreement on 
how the experimental data should be analyzed so that  different methods yield significantly
different values. In particular,  the two most recent datasets for
the proton (\cite{Xiong:2019umf,Bernauer:2010:NewMainz}) yield different values depending on how the
earlier set is analyzed. They also stated that  the uncertainties of the values of $r_p$ obtained from scattering are
 greater than  an order of magnitude (\cite{Xiong:2019umf,Hayward:2018qij,Gao:2021sml}) 
larger
than those resulting from the measurement of the Lamb
shift in muonic hydrogen and muonic deuterium. 

CODATA's decision to exclude electron scattering from the latest adjustment is understandable.  However, the situation is not quite as dire as stated  in the previous  paragraph.  The data for $G_E$ from the two experiments~(\cite{Xiong:2019umf,Bernauer:2010:NewMainz}) agree in the region $0.1 \le Q^2 \le 0.5\,\rm fm^{-2}$ as shown in Fig.~(\ref{comp})~(\cite{Xiong:2020kds,Gao:2024kkg}) and the analysis of the lowest values of $Q^2$ of the Mainz experiment leads also to the small radius~(\cite{Alarcon:2020kcz}).
While the precision of electron scattering does not yet match that of hydrogen spectroscopy, it seems reasonable to suggest that the results of electron scattering  support the small value of the radius. This is because the PRad data taken at low values of $Q^2$  proved to be sufficient to determine the radius in a blind test done by the author and D.~Higinbotham.  In any case, the upcoming PRad-II experiment and others presented in Table 2 have the potential to match or approach the accuracy of the atomic results and settle any remaining differences.

\subsection{Beyond the Standard Model (BSM) Potential}

Now that the big radius {\it vs} small radius question has been settled, the probability  that direct comparisons between
$\mu^\pm-p$ and $e^\pm-p$ scattering in the  MUSE  experiment, could reveal violations of lepton universality has actually increased!  This is because the reported initial  4\% difference in the value of $r_p$ was too large to be accounted for in any BSM theory with an ultra-violet (well-defined at high energies) completion. Any remaining differences are at least 10 times smaller and are easier to accommodate in new theories.  The MUSE  experiment is running now and its results are eagerly anticipated. The ability to determine the effects of two-photon exchange are unique. Discovery of such effects would, in itself, be very important. Any presumably small difference remaining after the extraction of the two-photon effects could come from a violation of lepton universality.

\begin{ack}[Acknowledgments]

I thank J. Bernauer, H. Gao, R. Gilman, D. Higinbotham , J-P.  Karr,  M. Kohl and  E. Slate for useful discussions.

This work is partially funded   by the U. S. Department of Energy, Office of Science, Office of Nuclear Physics under Grant No. DE-SC0026252.
 \end{ack}


\bibliographystyle{Harvard}
\input{ProtonRadius.bbl}
\end{document}